\documentclass[tightenlines,aps,nofootinbib]{revtex4}

\usepackage{psfig,axodraw,float}
\usepackage{graphics}

\def\e{\epsilon}

\begin{document}

%\vspace{0.6cm}

\title{

\begin{flushright}
\vbox{
\begin{tabular}{l}
\small DESY 05-244\\
\small SFB/CPP-05-82\\
\   \end{tabular} }
\end{flushright}

A New Method for Calculating Differential Distributions Directly
in Mellin Space}

\author{
Alexander Mitov $\!\!$\thanks{ e-mail: alexander.mitov@desy.de}}
\affiliation{Deutsches Elektronensynchrotron DESY, Platanenallee
6, D-15735 Zeuthen, Germany}

\begin{abstract}

\vspace{2mm}

We present a new method for the calculation of differential
distributions directly in Mellin space without recourse to the
usual momentum-fraction (or $z$-) space. The method is completely
general and can be applied to any process. It is based on solving
the Integration-by-Parts identities when one of the powers of the
propagators is an abstract number. The method retains the full
dependence on the Mellin variable and can be implemented in any
program for solving the IBP identities based on algebraic
elimination, like Laporta. General features of the method are: 1)
faster reduction, 2) smaller number of master integrals compared
to the usual $z$-space approach and 3) the master integrals
satisfy difference instead of differential equations. This
approach generalizes previous results related to fully inclusive
observables like the recently calculated three-loop space-like
anomalous dimensions and coefficient functions in inclusive DIS to
more general processes requiring separate treatment of the various
physical cuts. Many possible applications of this method exist,
the most notable being the direct evaluation of the three-loop
time-like splitting functions in QCD.

\end{abstract}

\maketitle

\thispagestyle{empty}

\section{Introduction}

Achieving high precision in theoretical predictions is vital for
the success of present and future collider experimental programs,
as well as for the effective extraction of new physics from
experimental data. A significant part of the theoretical work
related to the experiment requires the evaluation of differential
distributions, with most current research efforts focusing on the
Next-to-Next-to-Leading Order (NNLO) or a higher level of
precision. Examples of such distributions are the fully inclusive
\cite{dis1,dis2,dis3,dis4,difeq,MVV,dism} and one-particle
inclusive \cite{disinc1,disinc2} DIS, the energy spectrum of
hadrons in $e^+e^-$ collisions \cite{ee1,ee2,eeNW}, the total
partonic cross-section \cite{HK,AM,AModd,RSN} and rapidity
distribution \cite{ADM} for Higgs and vector boson \cite{ADMPW}
production at hadron colliders, Drell-Yan \cite{DY1,DY2,ADMPDY},
transverse distribution of hadrons at hadron colliders
\cite{NDE1,NDE2, pp1,pp2,pp3} or particle spectra in the decays of
muon \cite{mu1} or heavy flavors \cite{b1,b2,b3,b4,t1,t2}. Another
important class of distributions that are universal and thus
underlay the description of many physical processes includes the
space- and time-like splitting functions
\cite{AP,CFP,MVVnf,spacelike1,spacelike2}, heavy flavor matching
conditions \cite{match} and the heavy quark perturbative
fragmentation function \cite{DiniMN,Dini1,Dini2}.

The various distributions can be classified according to the
number of kinematical variables they involve. Clearly, the larger
the number of variables, the more complicated the evaluation of a
distribution becomes.  In this paper we will restrict our
discussion to the case of distributions with a single kinematical
variable. This class of distributions involves many important
examples - some of them still significant open problems - like the
three-loop time-like splitting functions in QCD. The extension of
our discussion to cases with more than one variable will be rather
transparent.

The choice of the most efficient approach to the evaluation of a
particular single-scale distribution depends on its degree of
"inclusiveness". The fully inclusive observables, like the fully
inclusive coefficient functions in DIS \cite{MVV}, allow a
simplified treatment based on the optical theorem. This is however
a rare situation; most distributions of interest involve a
specific final state, which requires that all contributing
physical cuts of the relevant amplitudes be evaluated separately.

The purpose of this paper is to present a conceptually new
calculational method of general applicability. As will become
clear from the subsequent discussion, this method builds a bridge
between two very important and seemingly unrelated calculational
approaches as it provides a new perspective on the calculation of
single-scale distributions. Moreover, during all stages of
calculation this method requires no custom work and utilizes
tools, techniques and programs that are publicly available and
easy to implement in practice. Our method relies heavily on the
Integration by Parts (IBP) identities \cite{IBP}. It has the
important feature of being formulated in terms of variables that
are the most natural ones for the effective solving of the IBP
identities.

With the above-described applications in mind, let us properly
introduce the type of distributions $\sigma(z)$ that we will be
dealing with in this paper. Such distributions depend on a single
kinematical variable $z$. For example, $z$ can be the energy
fraction of a parton produced in $e^+e^-$ annihilation. We will
assume that this variable is conveniently normalized: $0\leq z\leq
1$. The distribution $\sigma$ is a scalar that is typically of the
following form:
\begin{equation}
\sigma(z) = \int {\rm dPS}^{(m)} \vert M(\{{\rm in}\}\to \{{\rm
out}\})\vert^2 \delta(z-f). \label{sigma}
\end{equation}

The factor ${\rm dPS}^{(m)}$ in Eq.(\ref{sigma}) is related to the
phase-space for the $m$-particle final state; it also contains the
measure for the virtual integrations (if present). The precise
form of this factor depends on the number of particles in the
initial state. For a single-particle initial state processes with
no virtual corrections it reads:
\begin{eqnarray}
{\rm dPS}^{(m)}&=&(2\pi)^{d}\delta(p_{\rm in}-\sum p_{\rm
out})\prod_{i=1}^m \left[dq_i\right] , \nonumber\\
\left[dq_i\right]&=& {d^dq_i\over (2\pi)^{d-1}}
\delta(q_i^2-m_i^2) . \label{ps}
\end{eqnarray}
In the case of processes with two particles in the initial state,
${\rm dPS}^{(m)}$ has similar structure. It is detailed, for
example, in \cite{AM}.

Typically, expressions like Eq.(\ref{sigma}) are UV and infrared
divergent and in the following we assume that all divergences have
been properly regulated by means of dimensional regularization.
Besides $z$, the distribution $\sigma(z)$ can depend on other
parameters. Since their presence is irrelevant to our discussion,
we will assume in the following that these have some fixed values
and we will suppress them in our notations. The function $f$
appearing in the argument of the $\delta$-function in
Eq.(\ref{sigma}) is a dimensionless scalar. Its form is specific
for each particular observable.

As a typical example we will consider the evaluation of the single
particle inclusive cross-section for massless quark production in
the decay of a colorless particle $V\to q +X$ (see also the
appendix). Including the corrections up to next-to-leading order
in the strong coupling and working in terms of bare quantities
(i.e. no UV renormalization is performed) one has:
\begin{eqnarray}
{d\sigma \over dz} &=& \int\left[dp_q\right]\left[dp_{\bar
q}\right] \vert M^{(0)}(V\to q+{\bar q})\vert^2
\delta\left(z-{2 p_V.p_q\over p_V^2}\right)\nonumber\\
&+& \int \left[dp_q\right]\left[dp_{\bar q}\right]
{d^dp_{g^*}\over (2\pi)^{d}} \vert M^{(1)}(V\to q+{\bar
q})\vert^2\delta\left(z-{2 p_V.p_q\over p_V^2}\right)\nonumber\\
&+& \int \left[dp_q\right]\left[dp_{\bar
q}\right]\left[dp_g\right] \vert M^{(1)}(V\to q+{\bar
q}+g)\vert^2\delta\left(z-{2 p_V.p_q\over p_V^2}\right) .
\label{example}
\end{eqnarray}

In the example above, $\vert M^{(k)}\vert^2$ denotes the terms
proportional to $\alpha_S^k$ in the squared matrix element for the
process $V\to q +X$ (see Fig.1). Clearly, the first line in
Eq.(\ref{example}) corresponds to the tree-level (Born)
contribution while the second and the third lines respectively
contain the contributions from the virtual and real-gluon emission
corrections at order $\alpha_S$. On the above example,
Eq.(\ref{sigma}) stands for any one of the three lines in
Eq.(\ref{example}).

Perhaps the most elegant approach to date for the evaluation of
distributions of the type Eq.(\ref{sigma}) was proposed by
Anastasiou and Melnikov \cite{AM} and further elaborated upon in
\cite{ADM,ADMPW}. Let us recall the salient features of this
method. One uses the distributional identity:
\begin{eqnarray}
2\pi i \delta(x) = {1\over x+i\epsilon}-{1\over x-i\epsilon} ,
\nonumber
\end{eqnarray}
to formally replace all $\delta$-functions appearing in
Eq.(\ref{sigma}) with propagators coinciding with the arguments of
the $\delta$-functions, i.e. one introduces the invertible mapping
$\widehat{P}$ acting only on $\delta$-functions:
\begin{eqnarray}
\widehat{P}\left[c \prod_i\delta(x_i)\right] = c \prod_i {1\over
x_i} , \label{map}
\end{eqnarray}
with $c$ an arbitrary function of the propagators. The utility of
the mapping (\ref{map}) is that it allows one to treat the object
on the right-hand side of Eq.(\ref{map}) with the usual IBP
identities \cite{IBP}. By solving these identities one reduces the
initial distribution $\sigma(z)$ to a combination of a small
number of irreducible objects. Eventually, one performs the
inverse mapping $\widehat{P}^{-1}$, thus expressing $\sigma(z)$ as
a linear combination (with simple known coefficients) of a small
number of well defined master integrals. From the IBP identities
it also follows that the master integrals satisfy a system of
differential equations, which can be solved to obtain their
$z$-dependence. To fully specify the solutions of the differential
equations, one has to prescribe the corresponding boundary
conditions; these can be extracted from an explicit evaluation of
the master integrals in a particular kinematical point like $z=1$.

In the phenomenological applications one also needs the Mellin
transform
\footnote{Note that usually the Mellin transform is defined
through the variable $N=n+1$, $N\geq 1$.}
of the distribution in question:
\begin{equation}
\sigma(n) = \int_0^1 dz z^{n}\sigma(z). \label{mellin}
\end{equation}

Performing the Mellin transform results in the evaluation of
integrals over, typically, combinations of polylogarithms and
rational functions of $z$. At present, and certainly for the case
of massless distributions, there exists a very good understanding
of the mapping between the classes of basic functions in $z$ and
$n$ spaces \cite{hpl1,hpl2,hpl3, harmsum2, harmsum3,
Blumlein:2005jg}. In the following discussion we will consider the
knowledge of $\sigma(z)$ as equivalent to that of $\sigma(n)$ and
vise versa, i.e. we will tacitly assume that one can always
perform the needed Mellin or inverse Mellin transforms. That is
definitely true for the massless case. In more complicated
situations one may have to resort to numerical methods to perform
the inverse Mellin transform \cite{MellinNumVogt, MellinNumMP}. In
any case, we need not bother about that point here. We will
consider our problem as solved, as long as we know either
$\sigma(z)$ or $\sigma(n)$.

\section{The Method}

In the present paper we would like to advocate a new approach to
the evaluation of the distribution in Eq.(\ref{sigma}). It aims at
the direct evaluation of $\sigma(n)$ without calculating it first
in $z$ space as is done at present. Our proposal is to explore the
obvious possibility that one can integrate over $z$ before
performing the phase-space and/or virtual integrations:
\begin{eqnarray}
\sigma(n) &=& \int_0^1 dz z^{n}\sigma(z) \nonumber\\
&=& \int_0^1dz z^{n}\int dPS^{(m)} \vert M(\{{\rm in}\}\to
\{{\rm out}\})\vert^2 \delta(z-f)\nonumber\\
&=& \int dPS^{(m)} \vert M(\{{\rm in}\}\to \{{\rm out}\})\vert^2
{1\over \left(f\right)^{-n}}. \label{sigma-N}
\end{eqnarray}

We see that as a result of the interchange of the order of
integrations the invariant $f$ enters the integrals as a
propagator raised to power $-n$. That power, however, should be
treated as an abstract parameter that takes arbitrary and not
fixed integer values.

By applying the mapping Eq.(\ref{map}), and interchanging the
order of integration as in Eq.(\ref{sigma-N}), one can bring the
original problem of calculating $\sigma(z)$ to the following form:
\begin{eqnarray}
\widehat{P}\left[\sigma(n)\right] &=& \sum \int {\cal{D}}~{1\over
P_1^{a_1}\dots }{1\over x_1\dots }{1\over f^{-n+s}},
\label{typical-term}
\end{eqnarray}
where ${\cal{D}}$ represents the appropriate measure originating
from the real and/or virtual integrations, $P_i$ denote the
propagators originating from the evaluation of the amplitude and
$x_i$ are the arguments of all phase-space $\delta$-functions (if
present). The argument of the $\delta$-function that defines the
observed fraction $z$ is denoted by $f$ and the powers $a_i$ and
$s$ are some fixed integers. This way, we have effectively reduced
the problem of the calculation of the differential cross-section
$\sigma(z)$ to the problem of evaluation of functions of the
following general form:
\begin{eqnarray}
S(a_1,\dots , a_p) = \int {\cal{D}}~ {1\over P_1^{a_1}\dots
P_p^{a_p}}. \label{S}
\end{eqnarray}
The function $S$ appearing in Eq.(\ref{S}) can also depend on
other fixed parameters.

In principle, scalar terms like the one in Eq.(\ref{S}) can be
simplified to a minimal set of terms by applying the Integration
by Parts (IBP) identities \cite{IBP}. Until now, there has been no
effective method for solving the IBP identities when one (or more)
of the powers $a_1\dots a_p$ is an abstract parameter (however see
\cite{Tarasov}). In the following, we present one very efficient
approach for solving this problem.

\section{Method for solving the IBP Identities in presence of an abstract
power}

It is very well known (see for example the book of Smirnov
\cite{smirnov} for detailed introduction) that when applied to the
object $S$ in Eq.(\ref{S}), the IBP identities result in a system
of linear homogeneous equations with rational coefficients that
relate functions $S$ with arguments shifted by $\pm 1$ relative to
each other. If all $a_i$'s were fixed integers, then by
successively relating terms that differ with $\pm 1$ one can
eventually express the original function $S(a_1,\dots , a_p)$
through a linear combination of several, say $m$, master integrals
$S_1(\{i_1\}),\dots , S_m(\{i_m\})$. The masters $S_j$ are special
cases of $S(a_1,\dots , a_p)$ with their arguments $(\{i_j\})$
taking special values. At present, the most popular method for
solving the IBP identities is the one of Laporta \cite{Laporta}.
It is based on solving the systems of linear homogeneous equations
directly, through Gauss elimination.

Clearly, if one of the parameters $a_i$ is not an integer this
procedure cannot work, since: 1) with only integer steps one
cannot relate the initial non-integer parameter to an element
$S(a_1,\dots , a_p)$ with only integer $a_i$'s and 2) the number
of steps in the Gauss elimination cannot even be specified when
one of the parameters is an abstract number.

In the following, we present a solution to this problem. In order
to facilitate our discussion we shall assume that $a_1,\dots
,a_{p-1}$ are integers having some specific values, while the last
argument, $a_p$, is an abstract parameter.

From Eq.(\ref{sigma}) it is clear that $\sigma(n)$ is a sum of a
number of terms of the type in Eq.(\ref{S}) that have different
values of their indexes $(a_1,\dots , a_p)$. It is very important
to observe, however, that the difference between any two values
that the index $a_p$ can take is always {\it an integer}, i.e.
$a_p-a_p'\in \mathbf{N}$.

This is a crucial observation, which one can use to modify the
strategy for solving the IBP identities in the following way.
First, one relaxes the requirement that the masters must have
integer-valued indexes. Second, as we will explain in a moment,
one can choose all masters in such a way that they all have the
same value, say $a_p=r \notin \mathbf{N}$, of their last index
i.e. the masters are all of the form:
$$ S(j_1,\dots , j_{p-1},r) = \int {\cal{D}}~
{1\over P_1^{j_1}\dots P_{p-1}^{a_{p-1}}} {1\over f^{r}},
$$
with the same $r$, and the $j_i$'s being fixed integers specific
to each master.

It is indeed possible to arrange that all masters have the same
value of the non-integer-valued index $a_p$. That follows from the
arbitrariness of the value of this parameter (we only assume that
it is non-negative). Since there is no preferred value for that
index, the IBP system has a sort of translational invariance along
the index $a_p$. One can understand this by saying that $r$ and
$r+k$, where $k$ is a fixed integer, are equally arbitrary.
Therefore, we can take as a reference value for the index $a_p$
the number $r$ which we will consider abstract but having fixed
value. Having done that, the ``translational" invariance along the
values of $a_p$ is now ``broken". Clearly, the value $r$ now plays
the role of a zero reference point much like the value $a_p=0$ in
the usual case when all indexes take integer values. Therefore all
one need to do is to measure in {\it integer units} how much the
value of the last index of an element $S$ is displaced from the
reference point $r$.

Next, we give a practical recipe of how to implement the above
idea. Let us work with the functions
\footnote{These are essentially the same as the functions $S$
introduced above; the difference is in the notation used for the
last argument.}
$B$:
\begin{eqnarray}
B(a_1,\dots ,a_{p-1},k) = \int {\cal{D}}~ {1\over P^{a_1}\dots
P_{p-1}^{a_{p-1}}}{1\over f^{-n+k}},\label{B}
\end{eqnarray}
where as the ``reference" point for the last index we take the
Mellin variable $n$.

As follows from Eqns.(\ref{sigma-N}) and (\ref{typical-term}) the
distribution $\sigma(n)$ takes the following form:
\begin{eqnarray}
\sigma(n) = \sum_{a_1,\dots ,a_{p-1},k} c_{a_1,\dots ,a_{p-1}}
B(a_1,\dots ,a_{p-1},k),\label{sigma-work-form}
\end{eqnarray}
where $c_{a_1,\dots ,a_{p-1}}$ are some known coefficients. To
construct the needed algebraic reductions, one first applies the
IBP identities on a generic monomial of the form:
\begin{eqnarray}
{1\over P^{\nu_1}\dots P_{p-1}^{\nu_{p-1}}f^{\nu_p}},\label{mon}
\end{eqnarray}
where all powers $\nu_i$ are treated as arbitrary parameters.
Next, one identifies each term of the form (\ref{mon}) appearing
in the IBP equations, with the function $B(\nu_1,\dots
,\nu_{p-1},\nu_p+n)$, followed by the substitution
$\nu_p\to\nu_p-n$. After this manipulation the Mellin variable $n$
is explicitly present as a parameter in the resulting equations.
They can be solved in any approach available, including the one of
Laporta.

A word of caution: one has to keep in mind that, as follows from
Eq.(\ref{B}), the functions $B$ implicitly depend on $n$.
Therefore, one should not confuse the integer value $k$ in the
last argument of the function $B$ with the absolute power of the
corresponding propagator $1/f$, but should think of it as the
``distance" - in integer units - from the reference power $n$.

Finally, one can map all integrals appearing in
Eq.(\ref{sigma-work-form}) to the masters obtained from the
solving of the just-described reduction. This mapping is done in
the standard way.

Next, we explain how one can extract the $n$ dependence of the
master integrals. Assume that an element $B(b_1,\dots ,b_{p-1},0)$
is a master integral (with $b_1,\dots ,b_{p-1}$ some fixed
integers). One can inspect the already solved IBP reduction and
read off from there the result for the element $B(b_1,\dots
,b_{p-1},-1)$. Note that this element differs from the master
$B(b_1,\dots ,b_{p-1},0)$ only by the value of the last index. If
the element $B(b_1,\dots ,b_{p-1},-1)$ is not a masters itself,
then it must be a linear combination of the master integrals:
\begin{eqnarray}
B(b_1,\dots ,b_{p-1},-1) = c(n) B(b_1,\dots ,b_{p-1},0) +G(n).
\label{difeq}
\end{eqnarray}
Here $c(n)$ is a known, typically not very complicated function,
and the term $G(n)$ is a homogeneous linear combination of all
master integrals, except for the master $B(b_1,\dots ,b_{p-1},0)$.
Eq.(\ref{difeq}) is a first order non-homogeneous difference
equation of the type $F(n+1) = c(n) F(n) + G(n)$ (recall
Eq.(\ref{B})) for the master $B(b_1,\dots ,b_{p-1},0)$. Clearly,
repeating this procedure for each one of the master integrals
found in the reduction run, one can derive a complete system of
difference equations for all the masters. Typically, one observes
certain hierarchy among the master integrals; the simplest ones
satisfy homogeneous equations (i.e. $G(n)=0$) that can be solved
in terms of $\Gamma$-functions. These integrals then comprise the
non-homogeneous terms for the equations of other masters, and so
on.

In case the element $B(b_1,\dots ,b_{p-1},-1)$ is also a master
integral, one should read off from the reduction the result for
the yet higher term $B(b_1,\dots ,b_{p-1},-2)$. One should
continue doing this until one reaches an element $B(b_1,\dots
,b_{p-1},-k)$ which is not a master itself but all elements
$B(b_1,\dots ,b_{p-1},-s)$ with $0\leq s < k$ are masters. The
result from the reduction for the element $B(b_1,\dots
,b_{p-1},-k)$ represents a $k$-th order difference equation for
the master integral $B(b_1,\dots ,b_{p-1},0)$.

To solve the resulting difference equations one can make use of
existing techniques. Such equations were analyzed and successfully
solved in the course of the evaluation of the three-loop anomalous
dimensions in QCD \cite{spacelike1,spacelike2} and of the two
\cite{difeq} and three-loop \cite{MVV} coefficient functions in
DIS. In most cases of physical interest the resulting difference
equations can be solved after expansion in $\e$ in terms of
harmonic sums or their generalizations. In simpler cases, one can
even solve these equations in closed form in terms of
hypergeometric and/or $\Gamma$-functions. Difference equations as
a way of calculating master integrals were also used by Laporta
\cite{Laporta}.

Upon solving the system of difference equations for the master
integrals, one has achieved a complete extraction of the
dependence of the masters on the Mellin variable $n$. The only
remaining thing to do is to specify the initial conditions for the
solutions of the difference equations. Typically, that would be
the value of the masters for $n=0$. This is an important fact. It
implies that to completely specify the master, one need to only
evaluate integrals that are fully integrated over the available
phase-space. These are pure numbers that do not depend on the
kinematical variable $n$. On the conceptual level, this is placing
the evaluation of certain not-completely inclusive observables one
step closer to the very familiar fully-inclusive case where,
thanks to the optical theorem, one can significantly simplify the
calculations by not considering separately all possible physical
cuts.

Often, one can reduce the number of fixed-$n$ integrals that have
to be evaluated by hand. This follows from the property of the
fixed-$n$ reduction that not all integrals corresponding to
initial conditions for the $n$-dependent masters are actually
independent. To explore this fact one has to perform a separate
fixed-$n$ IBP reduction where $n=0$ is taken from the very
beginning. We have observed in simple one- and two-loop reductions
as well as in rather complicated three-loop cases that this
procedure indeed generates additional relations between the
integrals corresponding to the initial conditions of the master
integrals. One might wonder about the cost of such an additional
run. That, however, should be of no concern since the fixed-$n$
reduction is much simpler and faster than the general-$n$ run one
has to be able to perform anyway. The computer load pays off with
the elimination of many of the integrals that otherwise have to be
computed by hand. A fixed-$n$ reduction can also be used as a
cross-check of the general-$n$ calculation. This is similar to the
use of Mincer \cite{mincer} in the three-loop DIS calculations in
\cite{MVVnf,spacelike1,spacelike2}.

\section{Partial Fractioning}\label{parfrac}

Consider a case where during the evaluation of the amplitudes one
gets a propagator that is not linearly independent from the
constraint $f$ defining $z$ (see Eq.(\ref{typical-term})).
Clearly, that can happen in many ways and such linear dependence
might even involve a group of several propagators. However, to
simplify our point as much as possible, we will only consider a
simple situation. Consider the function:
\begin{eqnarray}
F(n) = \int {\cal{D}}\times(\dots)\times {1\over (1-f)}{1\over
f^n},\label{F}
\end{eqnarray}
where $\dots$ stay for powers of other possible propagators. If we
were to evaluate this integral in $z$-space we would first replace
$f$ everywhere with $z$, as is implied by the factor
$\delta(z-f)$. Then the factor $1/(1-f)$ becomes just the number
$1/(1-z)$ and drops out of the integral. However, when we work in
Mellin space the constraint $f=z$ cannot be used anymore. If $n$
were some fixed integer, we could have applied partial fractioning
$n$-times and split the linearly dependent propagators $1/f$ and
$1/(1-f)$. For symbolic $n$, however, that cannot be done and one
should again resort to solving difference equations. This can be
done in the following way. The identity:
\begin{eqnarray}
{1\over (1-f)}{1\over f^{n+1}} = {1\over (1-f)}{1\over f^n} +
{1\over f^{n+1}},\nonumber
\end{eqnarray}
immediately translates into a difference equation for the function
appearing in Eq.(\ref{F}): $ F(n+1) = F(n)+G(n)$. This is a simple
difference equation with non-homogeneous part given by:
\begin{eqnarray}
G(n) = \int {\cal{D}}\times(\dots)\times {1\over f^{n+1}}.
\label{G}
\end{eqnarray}
The integral (\ref{G}) does not contain linearly dependent
propagators and can be evaluated by using the procedures described
previously.

Another way of eliminating the linear dependence among the
propagators is to expand the propagator $1/(1-f)$ in geometric
series. That would completely eliminate this propagator and one
would end up with a standard problem where the index $n$ is
replaced by $n+k,~k\geq 0$ (here one can apply the usual reduction
since the index $n+k$ is as arbitrary as the index $n$ is).
Finally, one would have to sum up the resulting expression over
the index $k$.

Our experience shows that the combination of the above methods is
sufficient to eliminate the appearance of linearly dependent
propagators in any situation.

\section{Concluding Remarks}

The method presented in this paper represents a conceptually new
approach for the evaluation of differential distributions. At the
same time it has the advantage that it shares features with other
available methods that previously were applied with impressive
success. We would like to critically compare our approach with
these methods and clarify its distinct applicability.

Our method has the following advantages over the direct momentum
space approach \cite{AM}. First, the solving of the IBP reductions
is much more efficient and fast. In practical terms that may not
be an issue in simple one- or two-loop cases, but we have checked
that it brings enormous improvement when applied, for example, at
three loops. Second, the number of master integrals being produced
in the course of solving the IBP's is smaller and, third, our
method only involves computation of integrals that are pure
numbers. This is a significantly easier task compared to the
calculation of the $z$-dependent initial conditions at fixed
values of $z$.

We also expect our method to bring new insights into the
calculations of distributions with more than one scale.

Our approach has a few common features with the method used in the
inclusive DIS calculations at two- and three-loops
\cite{KK,MVVnf,difeq, spacelike1,spacelike2,MVV}. Still our method
has wider, in fact completely general, applicability and it relies
for the formulation and solving of the IBP identities only on well
established, multipurpose and publicly available methods and
software. An important shared feature is the fact that the master
integrals in both methods satisfy difference equations in terms of
the Mellin variable $n$. In this respect our method benefits
greatly from these existing developments, since many of the
technical tools needed for its practical realization have been
already established. That includes the methods for solving the
difference equations, the current good understanding of the
appropriate functional bases in $n$ and $z$ spaces and their
tabulation. Finally, well established multipurpose software like
FORM \cite{FORM}, Summer \cite{Summer} and XSummer \cite{XSummer}
exist and are publicly available. They are capable of effectively
dealing with the necessary algebraical manipulations. In fact, our
method opens up new venues for the application of the techniques
from the two- and three-loop inclusive DIS.

As a first application of our method we have rederived the NLO
coefficient functions in $e^+e^-$ \cite{ee1} (in fact to all
orders in $\e$).

We also have experience with reductions at three-loops, where the
advantage of our method becomes apparent. Our method ensures
better performance of the IBP solving software and produces
smaller number of master integrals.

{\bf Acknowledgments.}

I would like to thank Sven Moch for many useful conversations and
the careful reading of the manuscript. I would also like to thank
Lance Dixon for related collaboration and Kirill Melnikov for
insightful discussions.

The research of A.M. is supported in part by the DOE under
contract DE-FG03-94ER-40833, the  Outstanding Junior Investigator
Award DE-FG03-94ER-40833 and by the start up funds of the
University of Hawaii. It was also supported in part by the
Deutsche Forschungsgemeinschaft in
Sonderforschungsbereich/Transregio 9.

\appendix\section{Simple Example}

We present one example which is simple yet it demonstrates all
non-trivial features of the method discussed above. We consider
the evaluation of the coefficient function in the decay of a
colorless object to a quark-antiquark pair $V\to q+X$ at order
$\alpha_S$. One constructs $\vert M\vert^2$ in the usual way; the
relevant diagrams are shown on Fig.1. At leading order, the
differential observable of interest is:
\begin{eqnarray}
\sigma(z) = {1\over \sigma^{LO}}{d\sigma\over dz} = \delta(1-z) +
{\cal{O}}(\alpha_S).\label{exampleA}
\end{eqnarray}
Note that we have chosen normalization where the coefficient of
the $\delta$-function in the leading term is exactly one to all
orders in $\e,~ d=4-2\e$. In Mellin space, this corresponds to
$\sigma(n)=1$. The virtual corrections at order $\alpha_S$ produce
the same type of contributions.

Through order $\alpha_S$, all non-trivial $n$-dependence of the
distribution Eq.(\ref{exampleA}) originates from the real gluon
radiation diagrams. To be specific, we are interested in observing
the final state massless quark in the reaction $V(p)\to q(p-q-k)
+X$, where the unobserved massless antiquark and gluon carry
momenta $q$ and $k$ respectively. We take $p^2=1$. From the
independent momenta $p,q$ and $k$ one can construct five linearly
independent scalars that are needed to build the IBP reduction. As
such we choose:
$P_1=(p-q)^2,~P_2=q^2,~P_3=k^2,~P_4=(p-q-k)^2,~P_5=2-2p.q-2p.k$.

After we perform the mapping (\ref{map}) we have to deal with
monomials of the type:
\begin{eqnarray}
{1\over P_1^{a_1}}{1\over P_2^{a_2}}{1\over P_3^{a_3}} {1\over
P_4^{a_4}}{1\over P_5^{a_5}}\label{mon-ex}
\end{eqnarray}

Performing the IBP reductions in the Laporta's method implemented
in the program AIR \cite{AIR}, we find that there is a single
master integral:
\begin{eqnarray}
B(0,1,1,1,0) = \int d^dq_1 d^dq_2 \delta(P_1)\delta(P_2)
\delta(P_3){1\over P_5^{-n}}\label{B-ex}
\end{eqnarray}

By inspecting the results for the element $B(0,1,1,1,-1)$ from the
solved IBP reduction, we derive the following difference equation
for the only master integral:
\begin{eqnarray}
B(0,1,1,1,-1) = {2+n-2\epsilon \over 3+n-3\epsilon}~
B(0,1,1,1,0).\nonumber
\end{eqnarray}
It is trivial to solve this recurrence relation (we modify the
notation in an obvious way to make completely transparent the
$n$-dependence):
\begin{eqnarray}
B(0,1,1,1,0)(n) = {\Gamma(2+n-2\e)\Gamma(3-3\e)\over
\Gamma(3-3\e+n) \Gamma(2-2\e)}~B(0,1,1,1,0)(n=0) . \nonumber
\end{eqnarray}
The initial condition $B(0,1,1,1,0)(n=0)$ is defined through
Eq.(\ref{B-ex}) after setting $n=0$ there. It is a trivial to
compute number.

Our work is not quite done with the solving of the IBP reductions
and the evaluation of the master integral since there also appears
the propagator $P_{add}=(q+k)^2$ which does not belong to the set
$P_1,\dots ,P_5$. This propagator results from the right diagram
on Fig.1; it is not linearly independent from the set $P_1,\dots
,P_5$ but it can appears downstairs together with these
propagators. To resolve the situation one has to resort to the
partial fractioning technique discussed in section \ref{parfrac}.

Exploiting the constraints implied by the three $\delta$-functions
(with arguments $P_{2,3,4}$), one can easily establish that
$P_{add} = 1-P_5$. In this case one can apply the geometric series
trick discussed in the previous section to all terms where
$P_{add}$ is present downstairs:
\begin{eqnarray}
{1\over P_{add}}P_5^n = {1\over 1-P_5}P_5^n = \sum_{s=n}^\infty
P_5^s.\nonumber
\end{eqnarray}
Next one takes the summation outside the integrals; the resulting
integrand is of the type in Eq.(\ref{mon-ex}). With the help of
the IBP reduction this integral can be reduced to the master
integral discussed above. Inserting the explicit form of the
master, one obtains very simple expression containing only
$\Gamma$-functions. The summation over $s$ of this product of
$\Gamma$-functions can be easily performed and it results again in
a product of the same type of functions. Thus, the result from the
real emission radiation at order $\alpha_S$ can be easily
evaluated in closed form to all orders in $\e$. To use this
expression in practical applications one has to decompose it in
series in $\e$. This expansion can be easily automated with the
help of the programs Summer \cite{Summer} and XSummer
\cite{XSummer}.

\begin{figure}
\begin{center}
\begin{picture}(200,40)(0,-20)
\SetWidth{1}
\DashLine(-50,10)(0,10){3} \Vertex(0,10){2}
\ArrowLine(50,40)(0,10) \ArrowLine(0,10)(50,-20)
\Vertex(37.5,-12.5){2} \Gluon(50,15)(37.5,-12.5){2}{3}
\PText(-50,25)(0)[l]{p} \PText(60,40)(0)[l]{q}
\PText(60,-20)(0)[l]{p-q-k} \PText(60,15)(0)[l]{k}
\DashLine(150,10)(200,10){3} \Vertex(200,10){2}
\ArrowLine(250,40)(200,10) \ArrowLine(200,10)(250,-20)
\Vertex(237.5,32.5){2} \Gluon(250,5)(237.5,32.5){2}{3}
\PText(150,25)(0)[l]{p} \PText(260,40)(0)[l]{q}
\PText(260,-20)(0)[l]{p-q-k} \PText(260,5)(0)[l]{k}
\end{picture}
\end{center}
\caption{Real emission diagrams contributing to the decay of a
colorless particle $V(p) \to q(p-q-k) + X$ at ${\cal
O}(\alpha_s)$.}
\end{figure}
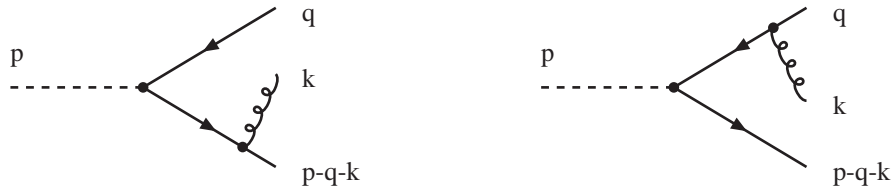
%
%

%%%%%%%%%%%%%%%%%%%%%%%%%%%%%%%%%%%%%%%%%%%%%%%%%%%%%%%%%%%%%%%%
%%%%%%%%%%%%%%%%%%%%%%%%%%%%%%%%%%%%%%%%%%%%%%%%%%%%%%%%%%%%%%%%

\end{document}